\documentclass[11pt]{article}
\usepackage{graphicx}
\usepackage{amssymb}
\usepackage{lineno}
\usepackage[titletoc]{appendix}
\usepackage{setspace}

\newif\ifFIGURE
\FIGUREtrue

\newcommand{\neffect}{$N^{\mathrm{eff}}_{\mathrm{pix}}$}
\newcommand{\nloca}{NLO$'_\mathrm{C\cdot A}$}

\title{\bf Describing the response of saturated SiPMs}

\author{Katsushige Kotera\footnote{Corresponding author ({\tt coterra@azusa.shinshu-u.ac.jp})}, Weonseok Choi\footnote{Current affiliation: {\it  Gumma R\&D Center, Advantest corp, 336-1 Owa, Meiwamachi, Ouragun, Gumma 370-0718, Japan }}, and Tohru Takeshita
\vspace{.3cm}\\
{\it Department of Physics, Shinshu University,}\\
{\it 3-1-1 Asahi, Matsumoto, Nagano 390-8621, Japan}}
\begin{document}
\maketitle
\baselineskip 1.0cm

\abstract{
We have developed a 
function which describes SiPM response in both small signal and highly saturated regimes.
The 
function includes the reactivation of SiPM pixels during a single 
input light pulse, 
and results in an approximately linear increase of
SiPM response in the highly saturated regime, as observed in real SiPMs.
This article shows that the 
function can accurately describe the measured response of real SiPM devices over a wide range of signal intensities.
}
\section{Introduction}
Pixelated photon detectors (PPD, also known as SiPM, MPPC) 
\cite{ANTICH1997, BONDARENKO2000, YAMAMOTO2007, Satoru} are becoming widely used in many fields which require compact photon detectors. 
These devices achieve a gain of $\mathcal{O}(10^5)$ with a bias voltage less than 100~V, and are insensitive to magnetic fields.
One complication is their intrinsically non-linear response, particularly at 
high input light levels, 
which can be described to first order by
\begin{equation}
N_{\mathrm{fire}}^{\mathrm{LO}} = N_{\mathrm{pix}}\Big( 1 - \mathrm{e}^{-\epsilon N_{\mathrm{in}}/N_{\mathrm{pix}}}\Big), 
\label{eq:first}
\end{equation}
in which $N_{\mathrm{fire}}$ is the number of fired pixels,
$N_{\mathrm{pix}}$ is the total number of pixels in the SiPM,
$N_{\mathrm{in}}$ is the number of photons that arrive at the sensor, and
$\epsilon$ is the single photon detection efficiency. 
%
This form\,--\,Leading Order (LO)\,--\,assumes that the charge produced in a pixel 
is the same if it is hit by one, two, or more photons.

However, when the {\it recovery time} of the SiPM pixels is shorter than the input light pulse 
({\it e.g.} Hamamatsu 25$\mu$m pitch MPPCs can have a  {\it pixel recovery time} $\sim$\,4\,ns.) 
a single pixel can contribute more than once to the output signal~\cite{Oide}.
H.~T.~van\,Dam~{\it et\,al.} \cite{Dam} have developed a model in which they implemented this pixel recovery as well as 
fluctuations of gain by charge release at pixels, dark noise, crosstalk, and after pulse. 
Dark noise is induced by thermal excitation, and 
the mechanisms of crosstalk  and after pulse are explained in section\,\ref{CrossAfterResult}. 
%
The model agrees within a few percent with the properties of SiPMs coupled to  $3\,\times\,3\times\,5$\,mm$^3$ LaBr$_3$:5\%Ce.
In their experiments,
the maximum number of fired pixels was $\sim$\,5 times of the number of real pixels. 
Their model is therefore rather complete. 
%
Their model is however mathematically rather complex, which limits its applicability to the task of calibrating and unfolding the response of SiPMs in a real experiment.

Instead, $N_\mathrm{pix}$ in Equation\,\ref{eq:first} can be replaced by a fitting parameter, \neffect, leading to what we call the LO$'$ description:
\begin{eqnarray}
N_{\mathrm{fire}}^{\mathrm{LO'}} = N_{\mathrm{pix}}^{\mathrm{eff}}\Big( 1 - \mathrm{e}^{-\epsilon N_{\mathrm{in}}/N_{\mathrm{pix}}^{\mathrm{eff}}}\Big)\label{eq:firstprime}.
\end{eqnarray}
This form has been used in {\em e.g.}~\cite{ScECAL-1}, using experimentally determined values of \neffect.
It does not, however, describe the linear increase in SiPM signal often seen at high light levels~\cite{linearPaper}.
In this paper we develop a 
function which can describe the SiPM response over its entire range, from small to highly over-saturated signals, 
using a function simple enough to use in SiPM calibrations.

The measured responses of 73 SiPMs are presented in the next section,
 and we develop functions to fit these data in section~\ref{DevelopFunctions}.
 We discuss the fitted parameters in section~\ref{Discussion}, and then conclude in section~\ref{Conclusion}.
 
\section{Experiment}\label{sec:data}

\begin{figure}[h!]
\ifFIGURE
\centerline{\includegraphics[width=0.4 \columnwidth]{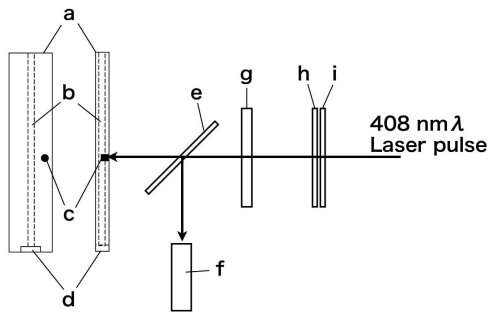}}
\fi
\caption{\small Setup of the $N_{\mathrm{pix}}^{\mathrm{eff}}$ measurement: a) target scintillator enveloped in reflector
(left, top view; right, side view);
b) WLS fiber; c) irradiation position with a small hole in reflector; d) MPPC; e) half mirror:
f) photomultiplier tube; g) lens; h) polaroid (fixed); and i) polaroid (rotatable).}\label{experiment}
\end{figure}

The 
responses of 72 1600-pixel MPPCs (with a pixel pitch of 25\,$\times$\,25\,$\mu$m$^2$)\footnote{
Model MPPC-11-025M 5887
provided  
in 2008
by Hamamatsu K.K. \cite{HPK} 
, corresponding to S10362-11-25P 
} 
were measured.
Each MPPC was attached to the end of a 1~mm diameter wavelength shifting (WLS) fiber (Y-11\footnote{Kuraray Co. Ltd. \cite{KURARAY}}), which was inserted into a
3\,$\times$\,10\,$\times$\,45\,mm$^3$ scintillator strip.
Figure~\ref{experiment} shows the setup of the measurement.
The scintillator strip enveloped in reflecting film, the WLS fiber, and the MPPC 
were 
developed for the CALICE scintillator strip 
electromagnetic calorimeter 
for an International Linear Collider experiment \cite{ScECAL-1, Granada}.
The scintillator strips were obtained by cutting 45\,mm lengths from long 3\,$\times$\,10\,mm$^2$ cross-section scintillator bars.
The scintillator bars were produced by extrusion with a central hole to house the WLS fiber \cite{KNU}. 
The WLS fiber has a 10\,ns decay time \cite{timeWLS}. 
Therefore, this system emphasizes the effect of multiple hits on a pixel.
To demonstrate the effect of the length of the light pulse, a scintillator strip without WLS fiber or hole was also tested. 

A 408\,nm laser\footnote{PiL040X (Head) + EIG2000DX (Controller) provided by Advanced Laser Diode System A.L.S. GmbH} 
pulse  with a FWHM duration of 31 ps was used to illuminate the scintillator strip via a 2.5\,mm diameter hole in the reflector film.
Crossed polaroid films were used to control the laser intensity, and a half mirror was used to independently monitor the light intensity.

A bias voltage 
3\,V  above the breakdown voltage was applied to the MPPCs.
The voltage of the output pulses from the MPPC was integrated over 150\,ns
using a charge-sensitive analog-to-digital converter (ADC)\footnote{Charge sensitive ADC, C009 provided by Hoshin electronics Co. Ltd.},
 synchronized with a signal from the laser pulser.
This charge measurement was then converted into the number of fired MPPC pixels by means of the previously estimated single pixel charge.

To 
measure the single pixel charge, output signals were amplified
\footnote{Amplifier-shaper-discriminator (ASD) IC and 16-ch ASD board for the ATLAS experiment \cite{ASD} }.
The factor to 
convert the amplified ADC counts to the un-amplified ADC was determined as the ratio of 
mean ADC counts at 
a certain laser light intensity, which 
was measured both with and without the amplifier.
Measurements were made at temperatures between 22$^{\circ}$C and 24$^{\circ}$C, 
and temperature-dependent effects on the gain of SiPM were corrected by using the previously measured linear relation between gain and temperature.

The intensity of laser light 
was simultaneously measured using a photo-multiplier tube (PMT)
\footnote{H1941-01c provided by Hamamatsu K.K.}
 whose response in the range of this experiment was confirmed to be linear. 
The integrated charge of the PMT pulse was measured using the same charge-sensitive ADC used to measure MPPC signals.
The ratio between the ADC counts of the PMT and the number of photons incident on the MPPC is 
 combined with the photon detection efficiency in the parameter $\epsilon$.

\begin{figure}[h!]
\ifFIGURE
\centerline{\includegraphics[width=0.5 \columnwidth]{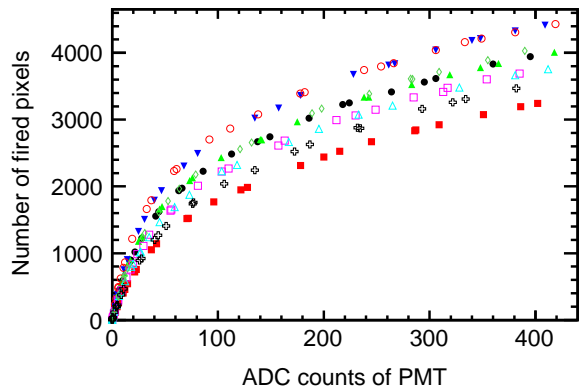}}
\fi
\caption{\small 
Responses of nine MPPCs.
The responses 
approach
linear behavior 
at 
incident light intensities 
over 
100 PMT ADC counts. 
}\label{ninechannels}
\end{figure}

Figure~\ref{ninechannels} shows the responses of a typical group of nine MPPCs as functions of the PMT signal. 

 \section{Developing the models and fitting to data}\label{DevelopFunctions}

 \subsection{Fitting with the LO$'$}
\begin{figure}[h!]
\ifFIGURE
\centerline{\includegraphics[width=0.5 \columnwidth]{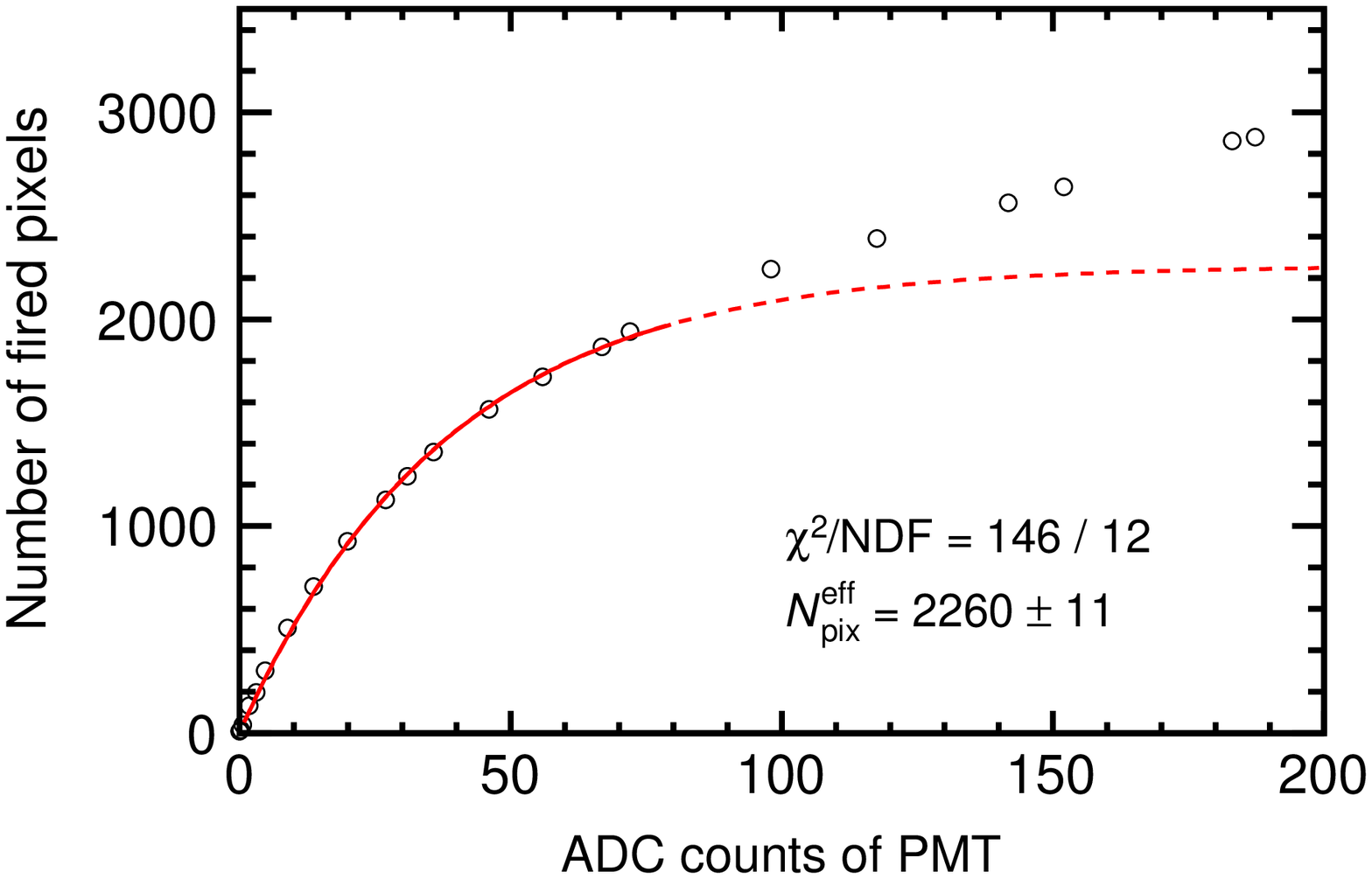}
\includegraphics[width=0.36 \columnwidth]{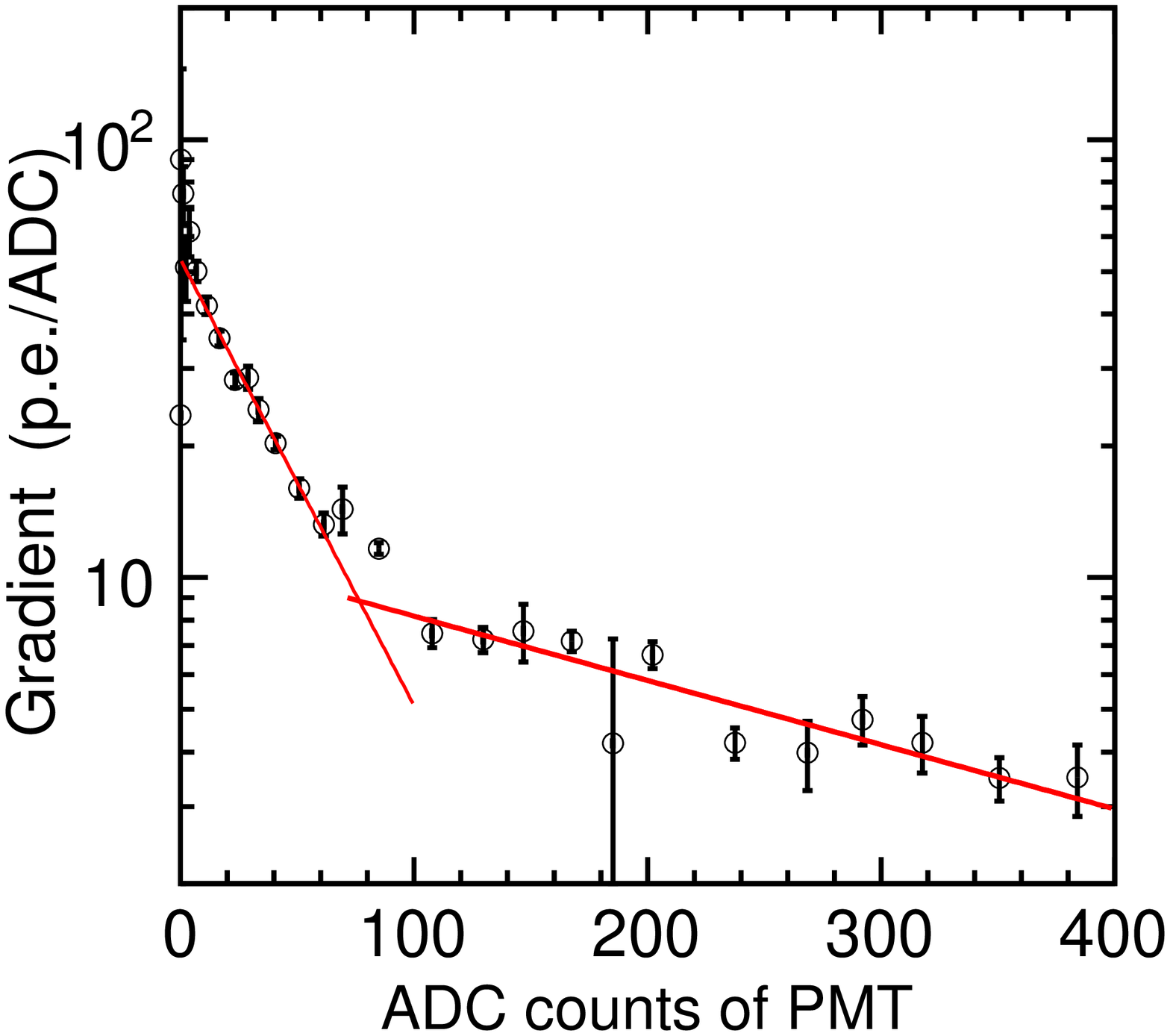}}
\fi
\caption{\small 
A fitting result of Equation\,\ref{eq:firstprime} (LO$'$) to MPPC response as a function of the incident photon measurement using PMT ({\it left}).
Range was determined from the intersection point of two fitting results on the difference of the $N_{\mathrm{fired}}$ with respect to the PMT response  ({\it right}).
}\label{fitLO}
\end{figure}

To represent the MPPC response shown in Fig.~\ref{ninechannels} with Equation\,\ref{eq:firstprime}, LO$'$, it is necessary to limit the range of the fitting.
Otherwise, the deterioration of fitting produces a meaningless result.
Figure~\ref{fitLO} {\it left} shows an example of fitting result with Equation\,\ref{eq:firstprime}, LO$'$ in a limited range; 
%
the inflection point of differential of the $N_{\mathrm{fired}}$ difference ($\Delta N_{\mathrm{fired}}) $ in a logarithmic scale with respect to the ADC counts of the PMT determines 
the upper limit of the fitting range. 
Figure~\ref{fitLO} {\it right} shows such a plot showing a kink of the linear behavior around 80 ADC counts.

The average of \neffect\ of 72 MPPCs is 2428 with a standard deviation of 245. 
This large \neffect\ indicates the reactivation of the pixels.
Although constant correction for the reactivation phenomenon such as \neffect\ works in the limited range shown in Fig.~\ref{fitLO},
rather linear like behavior is apparent after saturation.

\subsection{Next to leading order}
\begin{figure}[h!]
\ifFIGURE
\centerline{\includegraphics[width=0.4 \columnwidth]{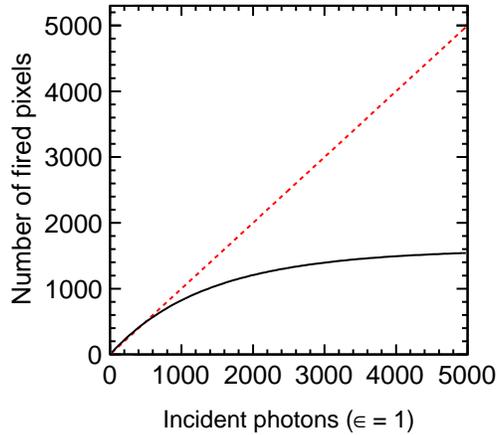}}
\fi
\caption{\small Ideal response of a 1600 pixel SiPM (solid curve) and response without saturation (dashed line) with parameter $\epsilon = 1$. 
}\label{modelPlot}.
\end{figure}

Figure \,\ref{modelPlot} shows the LO SiPM response (according to Equation\,\ref{eq:first})
as a function of the number of input photons, and a line representing unsaturated response.
%

The photons which do not fire a pixel at LO {\it i.e.} the number of which corresponds to the distance between the curve and line, can in 
principle contribute to multiple pixel hits with some probability.

Therefore, 
the number of remaining photons which could cause a multiple hit 
becomes 
\begin{equation}\label{NLO}
N_{\mathrm{R}} = \epsilon N_{\mathrm{in}} - N_{\mathrm{fire}}^{\mathrm{LO}}.
\end{equation}

\begin{figure}[h!]
\ifFIGURE
\centerline{\includegraphics[width=0.5 \columnwidth]{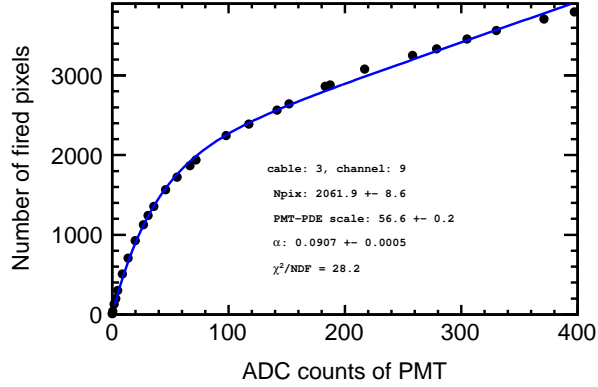}}
\fi
\caption{\small Example of fitting results of Equation\,\ref{eq:NLO} (NLO) to an MPPC responds as a function of the incident photon measured with PMT. 
Discrepancy still exists in the high light regime.
}\label{fig:fittingNLO}
\end{figure}

With a ratio of the average signal produced by each of these photons, $\alpha$ (this factor takes into account
effects due to {\em e.g.} different quantum efficiency and signal charge for later hits),
the total signal at next-to-leading order (NLO) becomes
\begin{eqnarray}\label{eq:NLO}
N_{\mathrm{fire}}^{\mathrm{NLO}} & = & N_{\mathrm{fire}}^{\mathrm{LO}} + \alpha N_{\mathrm{R}}. 
\end{eqnarray}
Figure \ref{fig:fittingNLO} shows a typical fitting result of Equation\,\ref{eq:NLO}.

Although a NLO represents the larger number of fired pixels than $N_{\mathrm{pix}}$  in the highly saturated regime, it does not represent the weak concave behavior
in this regime, and the fitting results of the number of pixels is $4/3$ times larger than 1600 pixels.
Therefore, we add additional corrections to NLO in the successive subsection.

\subsection{Charge contribution of a photon as a function of the number of photons on a pixel}\label{sec:charge}
The charge contribution of one photon decreases as the number of photons on a pixel increases, because the allowed time for recovery for each photon decreases.
The following explains more quantitatively that the charge contribution of one photon is a 
function of the ratio of the number of incident photons, $\epsilon N_\mathrm{in}$, to LO, and 
we use this relation to correct NLO.

D.~Jeans developed a model for the SiPM response from a first principle method \cite{Daniel}.
The model considers that the charge produced by an avalanche induced 
within a SiPM pixel depends on the recharge time from preceding avalanche.
 It assumes that; 
 \begin{itemize}
 	\item  the {\it recovery time} is exponential with a decay time, $\tau_R=RC$,
where the $R$ is the quenching resistance and $C$ is the capacitance of the pixel, and
	\item the pulse shape coming into SiPM from the WLS fiber or scintillator also exponentially decays with $\tau_s$.
\end{itemize}
According to Equation (2.9) of \cite{Daniel} the average charge by single photon is the following:
%
\begin{equation}
Q_1^\mathrm{(k)} = Q_0 \Big[ 1 +  \sum_{j = 2}^{k}\Big\{ 1-\frac{\tau_R}{\tau_R + \tau_s/(k-j+1)} \Big\}\Big]k^{-1},  \label{eq:chargeQ}
\end{equation}
where  $k$ is the number of photons come on a single pixel, $Q_0$ is the charge produced by a fully-charged pixel, and the sum only enters if $k\ge2$.

\begin{figure}[h!]
\ifFIGURE
\centerline{	\includegraphics[width=0.4 \columnwidth]{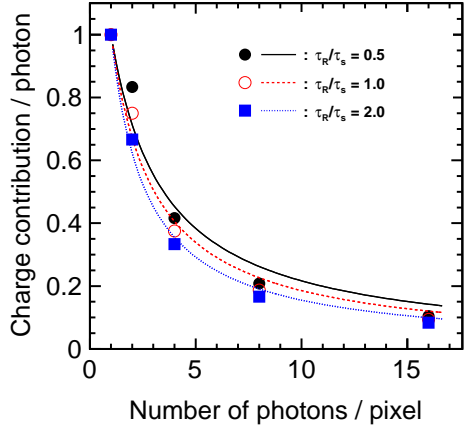}}
\fi
\caption{\small 
The charge contributions corresponding to single photon as a function of the number of photons on a pixel.
}\label{fig:charge_k}
\end{figure}

Figure \ref{fig:charge_k} shows $Q_1^{(k)}$ in Equation\,\ref{eq:chargeQ} as a function of $k$ for various $\tau_R/\tau_s$.
The curves show fitting results of 

\begin{equation}
{Q_1^\mathrm{(k)}}'= \frac{\beta +1}{\beta + k}, \label{eq:chargeQk}
\end{equation}
to the calculated points using Equation\,\ref{eq:chargeQ},
where $\beta$ is a fitting parameter. 
This simple Equation\,\ref{eq:chargeQk} approximately represents Equation \ref{eq:chargeQ}.
%

The average $k$ is the total incident photons $\epsilon N_\mathrm{in}$ divided by the net number of fired pixels of this SiPM, {\it i.e.} LO.
Therefore, the estimated fired pixels corrected for the effect of $k$ becomes the following:
\begin{equation}
N_\mathrm{fire}^\mathrm{NLO'} = N_\mathrm{fire}^\mathrm{NLO}\frac{\beta +1}{\beta + \epsilon N_\mathrm{in}/\mathrm{LO}}.\label{eq:NLOprime}
\end{equation}

\begin{figure}[h!]
\ifFIGURE
\centerline{	\includegraphics[width=0.5 \columnwidth]{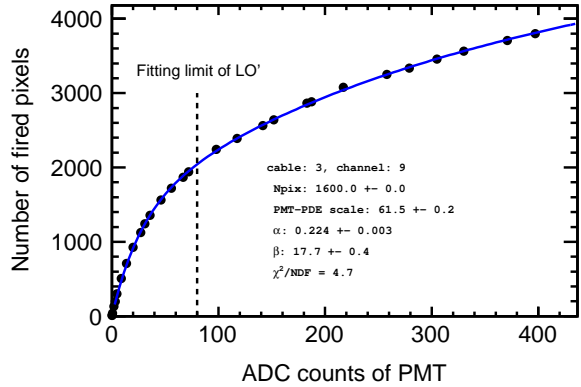}}
\fi
\caption{\small 
Example of fitting results of NLO$'$ {\it i.e.} Equation\,\ref{eq:NLOprime} to an MPPC responds as a function of the incident photon
measured with PMT. A vertical dashed line shows the fitting limit of LO$'$ {\it i.e.} Equation\,\ref{eq:firstprime}. }\label{fig:fit_fixNpix1600NLOprime}
\end{figure}

Figure \ref{fig:fit_fixNpix1600NLOprime} shows an example of fitting 
with 
Equation\,\ref{eq:NLOprime} (NLO$'$).
The discrepancy at high light regime in Fig.\,\ref{fig:fittingNLO} has disappeared.
The dashed vertical line is the fitting limit for LO$'$ showing that the applicable range is more than five times 
larger.
Two new parameters, $\alpha$ and $\beta$, are introduced in Equation\,\ref{eq:NLOprime}.
On the other hand the number of pixels, $N_\mathrm{pix}$ is fixed to 1600 pixels.
Therefore, only one increase of the number of  parameters 
improves the fitting performance.
Even if the fitting parameter  $N_\mathrm{pix}$ becomes free, the results of $N_\mathrm{pix} =1583\,\pm\,29$ 
is consistent with the true number of pixels.

\subsection{Crosstalk and after pulse}\label{CrossAfterResult}
Other considerable influences on the SiPM response are crosstalk and after pulse.
Crosstalk occurs when a photon
, created by the primary avalanche,
 subsequently induces a second avalanche in a neighboring
pixel \cite{Satoru}.
An after pulse occurs when a second avalanche is seeded by the release of 
an electron trapped in a lattice defect of the depletion zone,
or a hole 
defuses toward the depleted layer and
, consequently, induces a second avalanche.
The hole is created by a photon in the same mechanism as the crosstalk, and 
this diffusion delays the after pulses because the electric field of the bulk is weak \cite{afterpulsePD12}.
The resulting increase in the average number of fired pixels approximately becomes as
\begin{eqnarray}\label{eq:CrossAfter}
N^{\mathrm{NLO'_{C\cdot A}}}_{\mathrm{fire}}= N_{\mathrm{fire}}^{\mathrm{NLO'}}\Big( 1 + P_{\mathrm{cross}}\cdot \mathrm{e}^{- \epsilon N_{\mathrm{in}}/N_{\mathrm{pix}}}\Big)\cdot(1 + P_{\mathrm{after}}),
\end{eqnarray}
where $P_{\mathrm{cross}}$ represent the probability of crosstalk  and $P_{\mathrm{after}} $ represent the probability of after pulse per fired pixel. 
In this model, the crosstalk is proportional to the fired pixels, NLO$'$ and the ratio of unfired pixels to the total number of SiPM pixels, $\mathrm{e}^{- \epsilon N_{\mathrm{in}}/N_{\mathrm{pix}}}$,
because this immediate avalanche can only 
occurs 
on an unfired pixel.
The effect of after pulses is simply increasing the 
number of fired pixels in the first order.

\begin{figure}[h!]
\ifFIGURE
\centerline{	\includegraphics[width=0.4 \columnwidth]{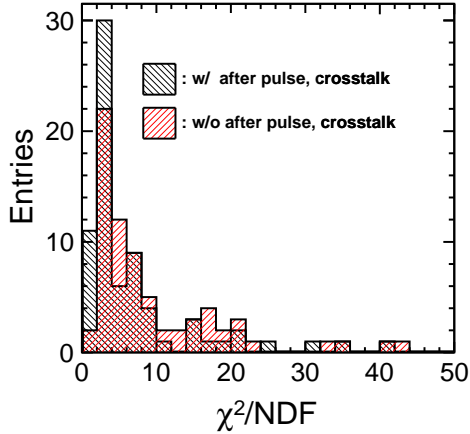}}
\fi
\caption{\small 
Distribution of the reduced $\chi^2$ of the fitting results of NLO$'$ and NLO$'_\mathrm{C\cdot A}$.}\label{fig:chisq_CrossAfter}
\end{figure}

Figure~\ref{fig:chisq_CrossAfter}  shows the distributions of reduced $\chi^2$ of fitting results with $N^{\mathrm{NLO'_{C\cdot A}}}_{\mathrm{fire}}$ (\nloca) 
and  NLO$'$.
The number of pixels is fixed to 1600 for both cases.
Two corrections for crosstalk and after pulse improve the fitting.
The mean of crosstalk probability is $0.22\,\pm\,0.02$ and the mean of after pulse ratio is $0.075\,\pm\,0.019$.
The  direct measurement of the crosstalk probability--\{(dark counts $>$ 2\,p.e.)\,/\,(dark counts $>$ 1\,p.e.)\}--with a 3\,V over voltage has given $0.11\,\pm\,0.02$ (RMS) for our MPPCs. 
The after pulse probability of the same type of MPPCs was measured as 10\% with a 3\,V over voltage \cite{Eckert}.

\subsection{Applying to fast light pluse}\label{fitting_NLOprime_plus}

\begin{figure}[h!]
\ifFIGURE
\centerline{	\includegraphics[width=0.5 \columnwidth]{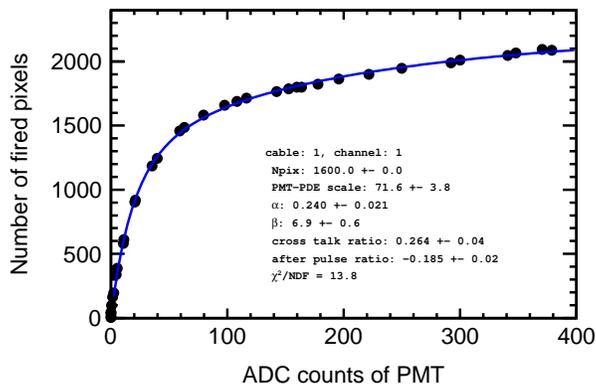}}
\fi
\caption{\small 
Example of fitting results of NLO$'_\mathrm{{C\cdot A}}$ to an MPPC 
response 
of the direct coupling MPPC-scintillator as a function of the incident photon measured with PMT.}\label{fit_fastpulse}
\end{figure}

Figure\,\ref{fit_fastpulse} shows a fitting result of \nloca\ to the response of MPPC coupled to a no-WLS fiber scintillator strip.
The fitting result indicates that \nloca\ is applicable to the response of MPPC-scintillator module having larger $\tau_R/\tau_s$  than  via the WLS fiber, {\it i.e.} long light pulse. 
The  reduced $\chi^2$, 14, is in the range of 72 samples with WLS fiber. 
The obtained parameter $P_\mathrm{cross} = 0.26\,\pm\,0.04$ is larger than our expected value, $\sim$\,0.1, and $P_\mathrm{after}$ is negative, those are still in one sigma of 72 data. The following section includes discussion 
of 
those fitting parameters.

\section{Discussion}\label{Discussion}

The finally obtained function, \nloca, has seven fitting parameters, 
some of which can 
be fixed and still give reasonable fittings.
Fixing 
$N_\mathrm{pix}$ to 1600 pixels provides reasonable fitting results as already mentioned in section \ref{sec:charge}, and obtained value of freed  $N_\mathrm{pix}$ agrees with 1600 pixels, whereas Equation\,\ref{eq:NLO}, NLO, still requires larger $N_\mathrm{pix}$, 2208\,$\pm$\,257 (RMS).
These facts indicate that the simple relation between $\epsilon N_\mathrm{in}$ and LO succeeds to represent the charge contribution of one photon via the number of photons 
in a 
single pixel.

The parameter $\epsilon$ includes the photon detection efficiency and the scale factor of PMT sensitivity to one photon.
The variation of $\epsilon$ in 72 samples is  25\% of mean value  57.2 with free crosstalk and after pulse and with fixed $N_\mathrm{pix}$ to 1600 pixels.

The recovery ratio, $\alpha$ is $0.295\,\pm\,0.04$ (RMS), and $\beta$ which moderates the influence of $k$  is $11.7\,\pm\,2.7$ (RMS) for 72 samples with 
$N^{\mathrm{NLO'_{C\cdot A}}}_{\mathrm{fire}}$.
These 
two parameters have the largest impact to determine the behavior of response in the high light regime.
Alpha varies to $0.24\,\pm\,0.02$ and $\beta$ varies to $6.9\,\pm\,0.6$  for a MPPC-scintillator direct coupling sample. 
The directions of  both changes agree with the difference of pulse decay time; 12\,ns \cite{timeWLS} for  the WLS fiber Y11  \cite{timeWLS} and 
2.5\,ns for SCSN38\footnote{pulse shape from SCSN38 corresponds to SCSN61 in their catalogue}  \cite{KURARAY},
because large $\tau_R/\tau_s$, {\it i.e.} without WLS fiber decreases the chance of recovery and increases the effect of $k$.
We however need to investigate more various types of SiPM and SiPM-scintillator combinations to extract precise  relations between those parameters and $\tau_R/\tau_s$.

The mean of the parameter, $P_\mathrm{cross} = 0.22$ for 72 samples is rather larger than our direct measurement of crosstalk, 0.11\,$\pm$\,0.02.
The value can be easily affected by other relations than crosstalk and after pulse, because the parameter 
affects weakly on the 
behavior of the function in plots.
The after pulse ratio, $P_\mathrm{after}$, fluctuates more, 
because it includes not only the after pulse effect but also the fluctuation of scale factor of vertical axis, {\it i.e.} the number of measured fired pixels.
Actually, RMS is 0.17 for the mean value 0.08, and 29\% of 72 samples have a negative value of  $P_\mathrm{after}$ which is never allowed for the real after pulse ratio.
The fitting result, $P_\mathrm{after} = -0.19$ for the MPPC direct coupling sample shown in section \ref{fitting_NLOprime_plus} is also considered as  such a case.
Those fluctuations of parameters imply the existence of some systematic uncertainties.
We can expect that adding such uncertainties decreases the reduced $\chi^2$ of this study close to unity, because 
the only statistical uncertainties of the measured number of fired pixels and the ADC counts of PMT were currently implemented on each data for the fitting.

 
\section{Conclusion}\label{Conclusion}

We have developed functions to represent SiPM response in a wide range considering pixel recovery.
In the case of NLO$'$,
the addition of a single fitting parameter with respect to LO$'$
increases the applicable range of the fitting function by more than a factor five.
These functions are an approximation of the more complete model of H.~T.~van\,Dam {\it et\,al.} \cite{Dam}.
Therefore, the parameters are influenced 
by 
not only the original 
physical meanings 
 of parameters but also 
 by some distortions 
 due to 
 the approximations.
Actually, we have seen that the crosstalk and after pulse probabilities have large fluctuations in NLO$'_\mathrm{C\cdot A}$, 
whereas their impacts on the shape of the function are small.
Nevertheless, the functions NLO$'$ and NLO$'_\mathrm{C\cdot A}$ are valuable for  practical uses such as detector calibrations for experiments,
because 
they
keep simultaneously both performance and simplicity.

\section*{Acknowledgments}
The authors would like to thank the CALICE collaboration. In particular, the CALICE-ASIA group 
provided a layer of the prototype of scintillator electromagnetic calorimeter.
In the group Daniel Jeans and Satoru Uozumi have contributed to crucial important discussions.
The authors extend special appreciation to Toru Iijima of Nagoya University, who allowed us to use the laser device. 
This work is supported in part by a Grant-in-aid for Specially Promoted Research: a Global Research and Development Program of a State-of-the-Art Detector System for the ILC' of the 
Japan Society for the Promotion of Science (JSPS, Grant No. 23000002).

 \end{document}